\documentclass{article}
\usepackage{amsmath,amssymb,epsfig,bbm,ifthen,capt-of,calc}

\newcommand{\tmem}[1]{{\em #1\/}}
\newcommand{\tmop}[1]{\ensuremath{\operatorname{#1}}}
\newcommand{\tmstrong}[1]{\textbf{#1}}
\newcommand{\tmtextit}[1]{{\itshape{#1}}}

\newtheorem{definition}{Definition}
\newtheorem{lemma}{Lemma}
\newtheorem{proposition}{Proposition}
\newcommand{\tmfloatcontents}{}
\newlength{\tmfloatwidth}
\newcommand{\tmfloat}[5]{
  \renewcommand{\tmfloatcontents}{#4}
  \setlength{\tmfloatwidth}{\widthof{\tmfloatcontents}+1in}
  \ifthenelse{\equal{#2}{small}}
    {\ifthenelse{\lengthtest{\tmfloatwidth > \linewidth}}
      {\setlength{\tmfloatwidth}{\linewidth}}{}}
    {\setlength{\tmfloatwidth}{\linewidth}}
  \begin{minipage}[#1]{\tmfloatwidth}
    \begin{center}
      \tmfloatcontents
      \captionof{#3}{#5}
    \end{center}
  \end{minipage}}

\begin{document}

\title{The extent of computation in Malament-Hogarth
spacetimes}
\author{P.D.Welch\\ School of Mathematics, University of Bristol,\\ Bristol, England,\\BS8 1TW\\
p.welch@bristol.ac.uk }\maketitle

\begin{abstract}

  We analyse the extent of possible computations following Hogarth
  {\cite{Ho94}} in Malament-Hogarth (MH) spacetimes, and Etesi and N\'emeti
  {\cite{EtNe02}} in the special subclass containing rotating Kerr black
  holes. \ {\cite{Ho94}} had shown that any arithmetic statement could be
  resolved in a suitable MH spacetime. {\cite{EtNe02}} had shown that some
  $\forall \exists$ relations on natural numbers which are neither universal
  nor co-universal, can be decided in Kerr spacetimes, and had asked
  specifically as to the extent of computational limits there. The purpose of
  this note is to address this question, and further show that MH spacetimes
  can compute far beyond the arithmetic: effectively Borel statements (so
  hyperarithmetic in second order number theory, or the structure of analysis)
  can likewise be resolved:
  
  {\tmstrong{Theorem A.}} {\tmem{If $H$ is any hyperarithmetic predicate on
  integers, then there is an MH spacetime in which any query $? n \in H ?$ can
  be computed.}}
  
  In one sense this is best possible, as there is an upper bound to
  computational ability in any spacetime which is thus a universal
  constant$\text{ }$of the space-time $\mathcal{M}$.
  
  {\tmstrong{Theorem C.}} {\tmem{Assuming the (modest and standard)
  requirement that space-time manifolds be paracompact and Hausdorff, for any
  MH spacetime $\mathcal{M}$ there will be a countable ordinal upper bound, $w
  ( \text{$\mathcal{M}$)}$, on the complexity of questions in the Borel
  hierarchy resolvable in it.}}
\end{abstract}

\section{Introduction}

Hogarth has shown that not only any universal statement, such as Goldbach's
Conjecture, but any arithmetical statement can be resolved in finite time, in
a suitable Malament-Hogarth (MH) space-time (these are defined in 1.1 below).

Our main observations are twofold: firstly that there is no reason for Hogarth
to stop at first order statements of arithmetic: effectively Borel statements
(so hyperarithmetic in second order number theory, or the structure of
analysis) can likewise be resolved (Section 2):

{\tmstrong{Theorem A.}} {\tmem{If $H$ is any hyperarithmetic predicate on
integers, then there is an MH in which any query $? n \in H ?$ can be
computed.}}

In fact for any transfinite Borel statement about the integers one can define
a MH spacetime in which the statement can be resolved.

In one sense this is best possible, as secondly, there is an upper bound to
computational ability in any spacetime, which is thus a universal
constant$\text{ }$of the space-time $\mathcal{M}$ (Section 3). The following
is merely an observation:

{\tmstrong{Theorem C.}} {\tmem{Assuming the (modest and standard) requirement
that space-time manifolds be paracompact and Hausdorff, for any MH spacetime
$\mathcal{M}$ there will be a countable ordinal upper bound, $w (
\text{$\mathcal{M}$)}$, on the complexity of predicates in the Borel hierarchy
resolvable in it.}}

Etesi and N\'emeti noted in {\cite{EtNe02}} that some $\forall \exists$
relations on natural numbers which are neither universal $(\forall)$ nor
co-universal $(\exists)$, can be decided in certain MH spacetimes
(instantiated by rotating Kerr black holes) and they ask specifically as to
the extent of computational limits in such spacetimes.

{\tmstrong{Theorem B.}} {\tmem{The relations $R \subseteq \mathbbm{N}$
computable in the spacetimes of {\cite{EtNe02}} form a subclass of the
$\Delta_2$-relations on $\mathbbm{N}$; this is a proper subclass if and only
if there is a fixed finite bound on the number of signals sent to the observer
on the finite length path.}}

This is treated in Sect 2.2.

\subsection{History and preliminaries} Pitowsky {\cite{Pi90}} gives an account
of an attempt to define spacetimes in which supertasks can almost be completed
- essentially they allow the result of infinitely many computations by one
observer $O_r$ (he used the, as then unsolved, example of Fermat's Last
Theorem) performed on their infinite ({\tmem{i.e. }}endless in proper time)
world line $\gamma_1$, to check whether there exists a triple of integers $x^k
+ y^k = z^k$ for some $k > 2$ as a counterexample to the Theorem or not. If a
counterexample was found a signal would be sent to another observer $O_p$
travelling along a world line $\gamma_2$. The difference being that the proper
time along $\gamma_2$ was finite, and thus $O_p$ could know the truth or
falsity of the Theorem in a (for them) finite time, depending on whether a
signal was received or not. As Earman and Norton {\cite{EaNo93}} mention,
there are problems with this account not least that along $\gamma_2$ $O_p$
must undergo unbounded acceleration.

Malament and Hogarth alighted upon a different spacetime example. The
following definition comes from {\cite{EaNo93}}:

\begin{definition}
  $\text{$\mathcal{M}$=} (M, g_{\tmop{ab}})$ is a {\tmem{Malament-Hogarth (MH)
  spacetime}} just in case there is a time-like half-curve $\gamma_1 \subset
  M$ and a point $p \in M$ such that $\int_{\gamma_1} d \tau = \infty$ and
  $\gamma_1 \subset I^- (p)$.
\end{definition}

(Here $\tau$ is proper time.{\footnote{We conform to the notation of Hawking
\& Ellis {\cite{HE}} and so $I^- (p)$ is the {\tmem{chronological past}} of
$p$: the set of all points $q$ from which a future-directed timelike curve
meets $p$. The spacetimes, all derived from Malament and Hogarth's ``toy
spacetime'', are differentible manifolds with a Lorentz metric
$g_{\tmop{ab}}$, and are time-oriented.}}) As they remark this makes no
reference to the word-line of a receiver of messages $O_p$ travelling along a
$\gamma_2$, but point out that there will be in any case such a
future-directed timelike curve $\gamma_2$ from a point $q \in I^- (p)$ to $p$
such that $\int_{\gamma_2 (q, p)} d \tau < \infty$, with $q$ chosen to lie in
the causal future of the past endpoint of $\gamma_2$. \ As Hogarth showed in
{\cite{Ho92}} such spacetimes are not globally hyperbolic, thus ruling out
many ``standard'' space-times (such as Minkowski space-time). Earman and
Norton's diagram of a ``toy MH space-time'' is Figure \ref{toy} below.
Hogarth's perhaps more succinct picture is on the right.\\

\tmfloat{h}{tiny}{figure}{\epsfig{file=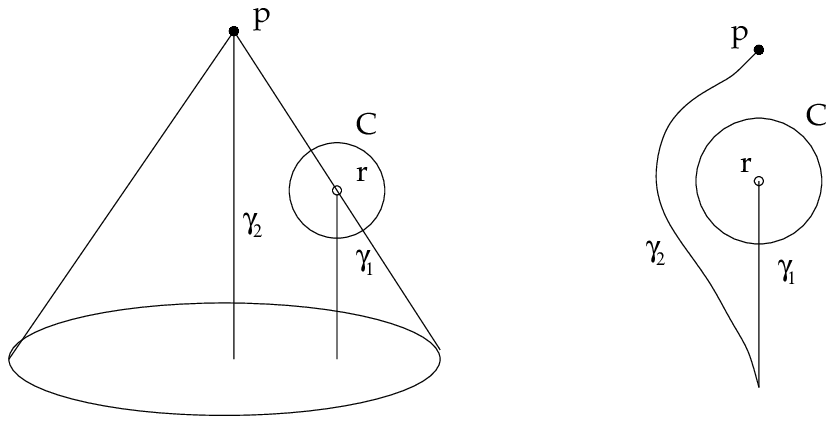}}{\label{toy}}\\

To rerun (and update) the Pitowski argument: the proper time along $\gamma_1$
is infinite, thus a Turing machine can be programmed to look for
counterexamples to the Goldbach Conjecture (or any other proposition involving
a single universal quantifier over an unquantified matrix: $\forall n P (n)$)
by checking $P (0), P (1), \ldots$ each of which in turn only takes a finite
amount of time; if a counterexample is found a signal is sent out to the
observer $\mathcal{O}_p$ travelling along the finite proper time curve
$\gamma_2$. If no signal is received by the time $\mathcal{O}_p$ reaches $p$
then $\mathcal{O}_p$ knows that the Conjecture is true. Either way
$\mathcal{O}_p$ is supposed to have discovered the truth by this point.

To obtain a spacetime as above, they take Minkowski spacetime $\mathcal{N}_0 =
(\mathbbm{R}^4, \eta_{\tmop{ab}})$ and choose a scalar field $\Omega$ which is
everywhere equal to 1 outside of a compact set $C$, and which rapidly goes to
$+ \infty$ as the point $r$ is approached. The point $r$ is removed and the MH
spacetime is then $\mathcal{N}= (\mathbbm{R}^4 \backslash \{r\},
g_{\tmop{ab}})$, where $g_{\tmop{ab}} = \Omega^2 \eta_{\tmop{ab}}$. $\Omega$
and $\gamma_1$ can be chosen so that $\gamma_1$ is a timelike geodesic.

Earman and Norton discuss at length the physical possibilities and
difficulties of this scenario: G\"odel spacetimes are MH, but are causally
vicious; the toy spacetime above need not satisfy any energy conditions;
anti-de Sitter spacetime is MH, but fails a strong energy condition;
Reissner-Nordstrom spacetime meets this, but as in all MH spacetimes there is
divergent blue-shift of the signal to $\mathcal{O}_p$; further, of the
unbounded amplification of signals that $\mathcal{O}_p$ may have to receive,
etc., etc. \ It is not our aim here to add to this discussion on
{\tmem{physical}} viabi{\tmem{}}lities but to make some observations on the
purely {\tmem{logico-mathematical}} possibilities and boundaries of this kind
of arrangement.

Logicians accordingly calibrate the complexity of propositions in the language
of arithmetic by the following hierarchy:

\begin{definition}
  (Arithmetical Hierarchy) Call a predicate $P \subseteq \mathbbm{N}^k$
  $\Sigma_0 $ or $\Pi_0$ if $P$ is recursive;
  
  $P \subseteq \mathbbm{N}^{k + 1}$ is in $\Sigma_n$ iff $\mathbbm{N}^{k + 1}
  \backslash P$ is in $\Pi_n$ ;
  
  if $\text{$P \subseteq \mathbbm{N}^{k + 1}$ is in $\Pi_n$}$ then $Q =\{(a_1,
  \ldots, a_k) | \exists b \in \mathbbm{N}(b, a_1, \ldots, a_k) \in P\}
  \subseteq \mathbbm{N}^k$ is in $\Sigma_{n + 1}$.
\end{definition}

A recursive predicate may be taken as one given by the extension of a
quantifier free formula of the language of arithmetic. (We sometimes identify
predicates, and their extensions, with their characteristic functions, and
shall implicitly assume in the future that for each $k$there is a recursive
bijection $r_k : \mathbbm{N}^k \longleftrightarrow \mathbbm{N}^{}$.) Deciding
whether $P (n)$ holds for a recursive predicate is then performable by a
computer or Turing machine in finite time.{\footnote{As explicitly here for
monadic predicates, but also {\tmem{via}} the coding functions $r_k$, for any
$k$-ary predicate also.}}

Hogarth in {\cite{Ho94}}, (and in the later {\cite{Ho04}}) uses the right hand
diagram of Fig.\ref{toy} above as a kind of short-hand for a component of
larger processes. He calls this a ``$\tmop{SAD}_1$ spacetime'' or region. A
segment of spacetime such as the above can be used to decide membership of any
$\Pi_1$ or $\Sigma_1$ definable set of integers (just as the set of even
integers which satisfy Goldbach's conjecture is a $\Sigma_0$ set, so the
statement of Goldbach's conjecture forms a $\Pi_1$ sentence, whose extension
is a truth value, {\tmem{i.e.}} is 0 or 1). If a spacetime contains a sequence
$\vec{O} = \langle O_j | j \geq 0 \rangle$ of non-intersecting open regions
such that (1) for all $j \geq 0$ $O_j \subseteq I^- (O_{j + 1})$ \ and (2)
there is a point $p \in M$ such that $\forall j \geq 0$ $O_j \subseteq I^-
(p)$ then $\vec{O}$ is said to form a {\tmem{past temporal string}} or just
{\tmem{string}}. To decide membership in a $\Pi_2$-definable set of integers
$P (n) \equiv \forall a \exists b Q (a, b, n)$ he then stacks up a string of
$\tmop{SAD}_1$ regions $O_j$ each looking like the component of Figure
\ref{toy}, with $O_0$ being used to decide $\exists b Q (0, b, n)$, if
this fails a signal is sent out to $\mathcal{O}_p$; but if this is successful.
a signal is sent to $O_1$ to start to decide $\exists b Q (1, b, n)$ etc.
Ultimately, putting this all together, again $\mathcal{O}_p$ receives a signal
if $\neg P (n)$, or else knows after a finite interval that it is true. It
should be noted that

{\tmstrong{Assumption 1}} {\tmem{The open regions $O_j$ are disjoint}},

{\noindent}and that no observer or part of the machinery of the system has to
send or receive infinitely many signals (``no swamping'' - we shall call this
{\tmstrong{Assumption 2}}). This whole region is then dubbed a
``$\tmop{SAD}_2$'' spacetime.

A ``$\tmop{SAD}_{n + 1}$'' spacetime is defined accordingly as composed from
an infinite string of (again disjoint) $\tmop{SAD}_n$ regions $O_n$, again all
in the past of some point $p$. (Earman and Norton {\cite{EaNo96}} show that a
$\tmop{SAD}_1$ spacetime cannot decide $\Pi_2$ statements. \ See the
discussion in Section 2.2 for what can occur if an arbitrary but finite number
of signals can be sent from the observer $\mathcal{O}_r$ to $p :$if the
spacetime is $\tmop{SAD}_1$ then $\Delta_2$ questions are resolvable. \
Hogarth {\cite{Ho04}} followed up {\cite{EaNo96}} with the generalisation that
$\tmop{SAD}_j$ cannot decide $\Pi_{j + 1}$ statements.) In the figure below,
if each $O_n$ is $\tmop{SAD}_j$ then the diagram on the left is that of an
$\tmop{SAD}_{j + 1}$ region. On the right is the underlying tree structure of
a $\tmop{SAD}_3$ region for computing queries of the form $? n \in A ?$ for
some $\Sigma_3$ set $A$.: each circle represents a $\tmop{SAD}_2$ region which
can be used for computing the answers to $\Sigma_2$ queries, and contains an
infinite string of $\tmop{SAD}_1$ components (pictured by the terminal nodes
of the tree) which appeared above at (the right hand side of) Fig. \ref{toy}.

\begin{figure}[h]
\quad\quad\quad\quad\quad\quad\quad
  \epsfig{file=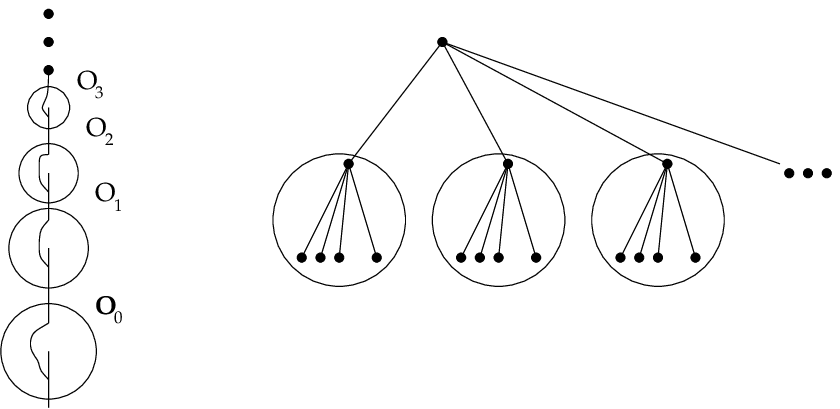}
  \caption{\label{sad3} A $\tmop{SAD}_3$ region as a past temporal string of
  $\tmop{SAD}_2 $regions and its tree representation (right).}
\end{figure}

\begin{definition}
  A spacetime $(M, g_{\tmop{ab}})$ is an {\tmem{arithmetic deciding (AD)
  spacetime}} just when it admits a past temporal string of disjoint open
  regions $\vec{O} = \langle O_j | j \geq 0 \rangle$ with each $O_j $a
  $\tmop{SAD}_{j + 1}$. 
\end{definition}

An AD spacetime is then a suitable manifold in which the truth of any
arithmetical sentence, or statement concerning any integer $n$, can be
determined. \ \ Again the various regions are connected in such an inductive
manner that no points receive infinitely many signals, and on {\cite{Ho94}} pp
131 following {\tmem{prima facie}} reasons of how the ``hardware'' fits in to
the regions concerned are given. An initial Turing machine is thought of as
``control'' of the process, in that, for example, it signals to activate the
appropriate $\tmop{SAD}_j$ region to compute the answer when given as input a
query whose complexity is $\Pi_j$. To actually define the AD spacetime, we may
adapt the toy spacetime example above so every component of our SAD regions
contains future directed half-lines approaching some ``removed'' point such as
$r$ above. Hogarth actually takes a closed inertial line segment $\nu$ inside
$C$, rather than a single point $r$; all the components of the $\tmop{SAD}$
regions are appropriately arranged around $\nu$ so that they intersect it. Now
$\Omega$ is chosen to rapidly approach $+ \infty$ as $\nu$ is approached, and
the segment $\nu$ is removed. The spacetime is then $\mathcal{M}_{\tmop{AD}} =
(\mathbbm{R}^4 \backslash \nu, \Omega^2 \eta_{\tmop{ab}})$ (see {\cite{Ho94}}
p133).

\section{Hyperarithmetic computations in MH spacetimes}

Our first remark is that this only scratches the surface of what is possible
in such spacetimes. The question of how far one can proceed in the
arithmetical hierarchy is also raised and discussed by Etesi \& N\'emeti in
{\cite{EtNe02}}. Our thesis is that one can decide questions far beyond
arithmetic in suitable spacetimes (Theorem A). \ However as we shall see,
under the mild assumptions that spacetimes are modelled by Hausdorff and
paracompact manifolds ({\cite{HE}} p.14), no one spacetime can decide all
Borel questions (Theorem B).

\subsection{Generalising $\tmop{SAD}_n$ regions}

First a definition: let $\widetilde{T} = \langle \widetilde{T},
<_{\widetilde{T}} \rangle = \langle \mathbbm{N}^{<\mathbbm{N}}, \supseteq
\rangle$ be the tree of all finite sequences of natural numbers ordered by
reverse inclusion (thus the empty sequence $\langle \rangle$ is the topmost
element of this tree and we view it as growing downwards as sequences are
extended). Members of $\widetilde{T}$ are thus finite functions: $v = \langle
v (0), v (1), \ldots, v (k) \rangle$ with each $v (i) \in \mathbbm{N}$.

\begin{definition}
  A {\tmem{finite path tree}} is any subtree $(T, <_T \rangle$ of
  $\widetilde{T} $where all branches under $<_T$ are of finite length.
\end{definition}

We assign {\tmem{ordinal ranks }}to the nodes of a finite path tree (which we
shall call just trees from now on) by induction: the terminal ``leaves'' at
the end of the branches are given rank zero, and in general the rank of a node
$u \in \mathbbm{N}^{<\mathbbm{N}}$ is the least strict upper bound of the
ranks of the nodes $v$which extend the sequence $u$. The {\tmem{rank of T}} is
then the rank of the empty sequence, $\langle \rangle$, the topmost node. The
point is that the tree, although all branches are of finite length, is in
general infinitely splitting (a given node $u$ may have infinitely many
immediate one step extensions); hence ranks of nodes can in general be
infinite, but of countable ordinal height (for an account of this and the
following context, see {\tmem{e.g. }}{\cite{Rog}} Sect.15.2).

We are less interested in the tree as being composed of finite sequences of
natural numbers, but rather the graph of the tree, or what it looks like: the
finite sequences simply give an informative way of indexing the nodes of the
tree. We shall be affixing other {\tmem{labels}} to the nodes of such trees.
Since an AD spacetime consists of an infinite string of regions $O_j$ where
each is an $\tmop{SAD}_{j + 1}$ spacetime region, which itself (if $j > 0)$ is
a string of $\tmop{SAD}_j$ spacetime regions, we can picture an AD spacetime
by the finite path tree:

\begin{figure}[h]
\quad\quad\quad
\hspace{3cm}  \epsfig{file=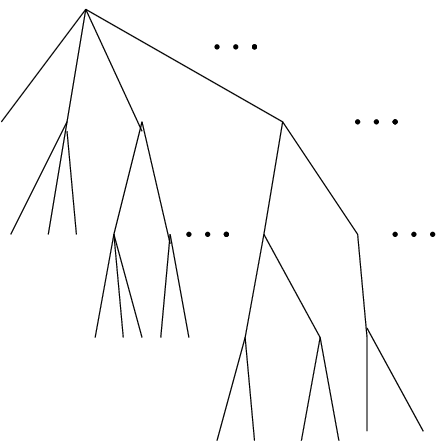}
  \caption{\label{AD} A finite path tree representation of an AD spacetime}
\end{figure}

Finite Path trees can be used to describe the construction of sets in the
{\tmem{Borel Hierarchy}} of sets of reals numbers. By ``the reals''  here we
mean infinite strings of integers from $\mathbbm{N}$, thus elements of
{\tmem{Baire space}} $\mathbbm{N}^{\mathbbm{N}}$ topologised by the product
topology on the discrete topology of $\mathbbm{N}$.{\footnote{ It is harmless,
and a commonplace practice, to identify the real line $\mathbbm{R}$ with the
set of such sequences - the irrationals are in any case homeomorphic to Baire
space, and omit only the countably many rationals. Nothing that we do, or
analytical problems that we attempt to solve in such spacetimes, depends on
any {\tmem{geometric}} consideration, or on the usual Euclidean metric on the
real continuum. \ }} A basic open neighbourhood in this topology is a set of
the form $N_s =_{\tmop{df}} \{x \in \mathbbm{N}^{\mathbbm{N}} | s \subset x\}$
where $s \in \mathbbm{N}^{<\mathbbm{N}}$ is a finite initial segment of $x$
(thus if $s$ is of length $k$, $s (i) = x (i)$ for $i < k) .$ We define in
marginally greater generality the Borel hierarchy in the space $\mathbbm{N}
\times \mathbbm{N}^{\mathbbm{N}}$ where $\mathbbm{N}$ has the discrete
topology and again the product topology is taken on this with the
$\mathbbm{N}^{\mathbbm{N}}$. A basic open neighbourhood here can be taken as
$N_s =_{\tmop{df}} \{(n, x) \in \mathbbm{N} \times \mathbbm{N}^{\mathbbm{N}} |
s (0) = n \wedge s (i + 1) = x (i) \tmop{for} i + 1 < \tmop{lh} (s)\}$

\begin{definition}
  (The Borel Hierarchy). (i) $X \subseteq \mathbbm{N} \times
  \mathbbm{N}^{\mathbbm{N}} $ is in $\Sigma_0$ and in $\Pi_0$ if it is a basic
  open set in the above topology; (ii) $X \in \Pi_{\xi} $iff $c X \in
  \Sigma_{\xi}$; (iii) $X \in \Sigma_{\xi}$ iff $X = \bigcup_n A_n$ where each
  $A_n \in \Pi_{\xi_n}$ for some $\xi_n < \xi$; a set $X$ is {\tmem{Borel}} if
  for some countable ordinal \ $\xi$ $X \in \Sigma_{\xi}$.
\end{definition}

(Here $c X$ is the complement of $X$.) It is well-known that such a hierarchy
is a) proper, that is $\Sigma_{\xi} \neq \Pi_{\xi}$ for $\xi > 0$, and with
both $\Sigma_{\xi}, \Pi_{\xi} \subsetneq \Sigma_{\xi + 1}$ for all $\xi <
\omega_1$, where the latter is the first uncountable cardinal number, and b)
the hierarchy terminates at $\omega_1 : \Sigma_{\omega_1 + 1} =
\Sigma_{\omega_1} .$ It is easy to reason that for any set $X \subseteq
\mathbbm{N} \times \mathbbm{N}^{\mathbbm{N}} $, if it is Borel, and let us
suppose that $\xi$ is least with $X \in \Sigma_{\xi}$, then, as per (iii),
there are infinitely many $A_n$ (possibly with repetition) so that $X =
\bigcup_n A_n$. We may then grow a finite path tree with $X$ at the topmost
node, then, infinitely splitting the tree below this node we have at the next
level down, the sets $A_0, A_1, \ldots$ which occur at lower levels in the
hierarchy. Each of these is in some $\Pi_{\xi_n}$ (with $\xi_n$ least) and
below the node for $A_n$ either (i) we place a single node labelled with $c
A_n$ if in fact $A_n \notin \Sigma_{\xi_n}$; or else (ii) $A_n \in
\Sigma_{\xi_n}$ and the tree infinitely further splits with nodes below that
of $A_n$ labelled for $B_m$ where $A_n = \bigcup_m B_m$ and each $B_m \in
\Pi_{\zeta_m}$ for some $\zeta_m < \xi_n$. In this way we can consider Borel
sets as built up according to a recipe, that can be described by a finite path
tree ({\tmem{finite}} because the ordinals are wellfounded) where the
determining labels are the (at most countably many) basic open sets attached
to each of the end nodes of each branch (and the appropriately constructed
sets labelling the nodes higher up, although of course the whole tree
labelling is determined by the tree structure and the assignment of basic open
sets at the terminal nodes).

As the reader can work out for his or herself, $\Sigma_{n + 1}$ predicates
$P$ of integers can be obtained by a process that can be given by a diagram
that has the structure of a tree of rank $2 n + 1$ (the $\Sigma_0$ sets are at
the bottom at rank 0, $\Sigma_1$ occupy the next layer up, $\Pi_1$ then occur
at the rank 2 next level after that; then come $\Sigma_2$ sets having rank 3,
and so on; the $\Sigma_{n + 1}$ set $P$ occupies the node labelled by the
empty sequence). The arithmetic sets are those that can be built up using
basic open sets in a finite number of stages: they are those sets of finite
rank: symbolically Arithmetic = $\bigcup_n \Sigma_n$.

Hogarth's AD spacetime is thus needed to calculate answers to membership
questions of the form $? n \in S$? where $S$ ranges over arithmetic sets. The
reader can probably now surmise what is going to happen: one can form sets of
integers $S \in \Sigma_{\omega}$ which are {\tmem{not}} arithmetic: they are
countable unions of such but there is no finite bound on their definitional
complexity. Such is represented by a tree with no finite bound on path length,
but has infinite rank. Actually one construal of an AD spacetime is that it
could answer questions of the form $? n \in S ?$ for {\tmem{some}} $S \in
\Sigma_{\omega}$: such an $S = \bigcup_m T_m$ where each $T_m$ is arithmetic.
If we assume (by expanding the list $T_n$ if need be with dummy sets $T_k =
\varnothing$) that each $T_m \in \Sigma_m$ we may query if $n \in T_m$ for
each $m$ in turn, where the $m$'th component of the AD spacetime is charged
with answering this question. Again if some $m$ is found for which $n \in T_m$
then a signal may be sent to the final observer... Notice that the tree
structure of Fig.\ref{AD} could equally well be used to describe the
components of the AD spacetime and the compositional structure of the set
$S$.{\footnote{Although the tree structure for an arithmetic set is slightly
expanded {\tmem{vis \`a vis}} the AD structure, as an $\tmop{SAD}_{j + 1}$
region can calculate truth values of both $\Sigma_j$ and $\Pi_j$ statements,
whereas drawing a tree for a $\Pi_j$ set adds one rank over an $\Sigma_j$ set,
but this is inessential difference which may be ignored.}} \

Why just {\tmem{some}} $\Sigma_{\omega}$ sets $S$ in the above? Because the
sequence of sets $\langle T_m | m \in \mathbbm{N} \rangle$ may not be
recursively, or {\tmem{effectively}} given to us. The description of this
sequence may itself be beyond the computational powers of us or our
spacetime-regions. Accordingly we first focus on a hierarchy of sets that can
be given an effective or algorithmic description. This special class of sets
of integers are wellknown and are called the {\tmem{hyperarithmetic}} sets
({\tmem{cf}}{\cite{Rog}} 16.8 or {\cite{Sa90}}): these are formed by protocols
that {\tmem{can}} be represented by finite path trees of {\tmem{recursive
ordinal}} height, where the terminal nodes are labelled with a recursively
given list of basic open sets.{\footnote{Recall that a recursive ordinal is
one for which there is a Turing machine $P_e$ which will compute a set of
integers $W$ that codes in some sensible fashion, a wellordering, $<_W$, of a
set of integers of length or {\tmem{order type}} that ordinal. There is a
least ordinal countable ordinal which is not so representable, and is known as
$\omega_1^{\tmop{ck}}$ (read ``Church-Kleene-omega-1''). }} Alternatively put,
the underlying recursive finite path trees $T$ are themselves the outputs of a
subset of the class of all Turing machines. (We thus imagine the machine
outputting an integer $k$ if, under some suitable computable coding, $k$codes
the fact that $u <_T v$ where the sequence $v$ is an initial segment of the
sequence $u$ in the tree $T$). In short, the hyperarithmetic sets are those
where there is a {\tmem{recursive description of their construction via finite
path trees}}. \ \

We may imagine then, in analogy to an AD spacetime, for a given
hyperarithmetic set $S$, a spacetime $\tmop{SAD}_S$ which consists of a
suitable piece of the manifold where there is a tree structure of
computational components reflecting the construction of $S$ from recursive (in
topological terms, basic open) sets. An initial ``control'' machine is given
$n$ (the integer on which the query is based), together with $e$ the code of
the machine which outputs the recursive tree structure $T_S$ of the set $S$
being queried. The spacetime has been chosen so that its open regions $O_f$
correspond to this recursive tree structure and so to the subqueries that
arise, and these will be arranged as nested subregions in a way reflecting
precisely the structure of $T_S$ (just as an $\tmop{SAD}_j$ region is
physically arranged or chosen so that it can answer $\Sigma_j$ or $\Pi_j$
queries). The adaptation of the toy spacetime construction for the case of an
AD spacetime, is the same as for any $\tmop{SAD}_S$ spacetime: it simply a
matter of ensuring the line segment $\nu$ intersects each of the regions of
the components appropriately.

We can be more specific in this regard: Hogarth needs a plentiful supply of
points where the metric goes rapidly to $+ \infty$. In defining a
$\tmop{SAD}_2$ region an $\omega$-sequence of points $\langle r_i |i < \omega
\rangle$ is needed which will eventually be removed from the manifold, but
towards each of which the metric goes to $+ \infty$ and there is some $q_i$
with $I^- (q_i)$ containing an endless half-line approaching $r_i$; for a
$\tmop{SAD}_3$ region each of the $\omega$-sequence of $\tmop{SAD}_1$ regions
is replaced by a $\tmop{SAD}_2$ region (which in turn contains an
$\omega$-sequence of removed points) ; thus a sequence of points of order type
$\omega^2 = \omega \times \omega$ where the metric goes to $+ \infty$ is
needed. Similarly a $\tmop{SAD}_{j + 1}$ region requires, according to the
Hogarth construction, an $\omega^j$ sequence of such points, and an
$\tmop{AD}$ region a $\omega^{\omega} (= \sup_j \omega^j)$ sequence. Now
suppose $S$ is hyperarithmetic and $\langle T, <_T \rangle$ is the recursive
subtree of $\langle \mathbbm{N}^{<\mathbbm{N}}, \supseteq \rangle$ whose
structure corresponds to the construction of $S$ once a recursive assignment
of recursive sets has been attached to the terminal nodes. We may embed the
tree in a (1-1)$<_T$-order preserving fashion, into the countable ordinals by
some function $\pi : T \longrightarrow \tmop{On}$ so that $u <_T v$ implies
$\pi (u) < \pi (v)$. We may do this by using the {\tmem{Kleene-Brouwer}}
ordering on $\langle \mathbbm{N}^{<\mathbbm{N}}, \supseteq \rangle$: $u
<_{\tmop{KB}} v$ if for the least $i$ so that $u (i) \neq v (i)$ either $v
(i)$ is undefined, or both $u (i), v (i)$ are defined and $u (i) < v (i)$. It
is routine to check that if $T$ is a finite path tree, then the Kleene-Brouwer
ordering restricted to $T$ is a wellordering, and hence is isomorphic to some
countable ordinal $\theta = \theta (T) < \omega_1$. We map each node $u \in T$
via $<_{\tmop{KB}} \upharpoonright T$ and the isomorphism just mentioned, to
an ordinal $\pi (u) < \theta$. We may then recreate a spacetime with the tree
structure of $T$ by using Hogarth's method on a sequence of points
$r_{\delta}$ $(\delta < \theta (T))$ which as approached, the metric will go
off to $+ \infty$. As \ any countable ordinal can be embedded into a (finite)
closed line segment this is unproblematic. This sequence of points will be
eventually removed. If $\delta = \pi (u)$ and $u$ has infinitely many
immediate tree $<_T$extensions $v$ of the form $v = u \frown \langle n
\rangle$, then $\delta$ will be a limit ordinal and the local region we may be
visualising, $O_u$, will contain at the next level down the infinitely many
subregions $O_v$ for $v$ of the above form. Formally then we may regard the
spacetime construction as one performed by induction on the $T$-rank of nodes
$u \in T$ when defining the regions $O_u$. We attach a Turing machine to all
nodes of rank 0, and if $\tmop{rank}_T (u) = \alpha$, then we gather all
immediate extensions of the form \ $v = u \frown \langle n \rangle ,$ (which
have rank $< \alpha)$ and connect these region-components $O_v$ according to
the KB-ordering on such $v :$in other words if $v_i = u \frown \langle n_i
\rangle$ for $i \in \omega$ enumerate such, then we ``connect'' \ the regions
so that $O_{v_i}$ lies in $I^- (q_{v_j})$ \ iff $v_i <_{\tmop{KB}} v_j$ (iff
$n_i < n_j$) just as in the left hand part of Fig. \ref{sad3}, with the
components approaching some point $r_u =_{\tmop{df}} r_{\pi (u)}$ with the
latter in $I^- (q_u)$ where $q_u =_{\tmop{df}} q_{\pi (u)}$ is some suitable
point. In other words given the finite path tree structure, we may define the
spacetime inductively from it. \ \ We formally state this argument as:

\begin{lemma}
  \label{induction}Let $(T, <_T)$ be a finite path of $\tmop{rank} \alpha$.
  Then there is a MH spacetime with regions $O_u$ for $u \in T$, with points
  $q_u$, so that
  
  (i) if $v <_T v$ then $O_v \subset O_u$ and  $O_v \subset I^- (q_u)$ ;
  
  (ii) if $v_i = u \frown \langle n_i \rangle$ $i \geq 0,$ enumerates the
  one-point extensions of $u$ with $u \frown \langle n_i \rangle \in T$, and
  $i < j \longrightarrow n_i < n_j$ then there is a past temporal string of
  regions $\vec{O} = \langle O_{v_j} | j \geq 0 \rangle$ with distinguished
  points $q_j \in O_{v_j}$, and $q_u \in O_u$, with \ $\{q_{v_i} \} \cup
  O_{v_i} \text{$\subset I^- (O_{v_j})$}$, and $O_{v_i} \text{$\subset I^-
  (q_u)$}$ connected as in Fig. \ref{sad3}. 
\end{lemma}

Proof: The Lemma may proven as an induction on $\alpha$. Let $v_i = \langle
n_i \rangle$ enumerate those one point extensions of $\langle \rangle$ with
$n_i$ in increasing order. For a given $i$ the restriction of $<_T$ to the
subtree of nodes in $T$ extending $v_i$, $T_{v_i}$, is a subtree of rank less
than $\alpha$. By induction we may construct suitable pieces of an MH
spacetime corresponding to those subtrees; call these $O_i = O_{v_i}$. As
Hogarth does for the AD spacetime, we may consider these constructions done so
that the ``removed points'' $r_{v_i}$ lie along some closed line segment $\nu$
where the metric wll go off to $+ \infty$, and then we string them together to
form a past temporal string. \ Alternatively the construction can proceed
directly again by induction but along the Kleene-Brouwer ordering restricted
to $T$. \ \ \ \ \ \ \ \ \ \ \ \ \ \ \ \ \ \ \ \ \ \ \ \ \ \ \ \ \ \ \ \ \ \ \
\ \ \ \ \ \ \ \ \ \ \ \ \ \ \ \ \ \ \ \ \ \ \ \ \ \ \ \ \ \ \ \ \ \ \ \ \ \ \
\ \ \ \ \ \ \ \ \ \ \ \ \ \ \ \ \ \ \ \ \ \ \ \ \ \ \ \ \ QED (Lemma
\ref{induction} and Theorem A)

Note, for a later discussion, that the construction does not {\tmem{require}}
that the finite path tree $T$be recursive: we may define the connections and
the regions by induction on the rank of nodes in any such $T.$

An alternative description of hyperarithmetic sets may prove helpful: suppose
$B \subseteq \mathbbm{N}$. Let $\chi_B \in 2^{\mathbbm{N}}$ be its
characteristic function. A {\tmem{code}} for $B$ is any function $f \in
\mathbbm{N}^{\mathbbm{N}}$ so that either:

(i) $f (0) = 0$ and $\forall n (f (n + 1) = \chi_B (n))$; or

(ii) $f (0) = 1$ and there is a code $g$ for $\mathbbm{N} \backslash B$, \ and
$\forall n (f (n + 1) = g (n)])$; or

(iii) $f (0) = 2$ and there is a sequence of sets $\langle C_k \rangle_{k \in
\mathbbm{N}}$ and a sequence of reals $f_k \in \mathbbm{N}^{\mathbbm{N}}$ so
that \ $f_k$ is a code for $C_k$, with $\forall n \forall k (f (r_2 (k, n) +
1) = f_k (n))$, and lastly $B = \bigcup_k C_k$.

By (i) trivially any set has a code! However the hyperarithmetic sets can be
characterised as precisely those sets $S$ which possess a recursive
code.{\footnote{Of course this is {\tmem{not}} to say that $S$ is itself
recursive, it is just that its {\tmem{construction}} has a recursive
description. The set of codes of hyperarithmetic sets is not recursive, or
r.e., or even arithmetic, it is complete $\Pi^1_1 $(see the next footnote).}}
If $S$ is hyperarithmetic we may think of the recursive code $f$ as given by
an index $e \in \mathbbm{N}$ for some Turing program $P_e .$ This latter
program computes $f (0)$ and, to take the more interesting case, if it is 2,
$P_e$ proceeds to compute values of the recursively given list of recursive
functions $f_k$ - which we may think of themselves as given by their index
numbers $e_k$. (Thus, {\tmem{inter alia}}, the sequence $\langle e_k
\rangle_{k \in \mathbbm{N}}$ is itself a recursive list.)

Hence from $e$ we may envisage a control program indexed by some $\tilde{e}$
which takes a query $? n \in S_e ?$and if $f (0) = 2$ it computes the
recursive list $\langle e_k \rangle_{k \in \mathbbm{N}}$ of indices of codes,
passing $n$ and these indices to each in turn of an infinite list of machines
at the next level down; these machines compute $f_k (0)$ in turn, and act
according to the $0, 1, 2$ outcome. Eventually, as the terminal leaves in the
tree are reached, the last code is that of a basic open set, and any such is
recursive. We conclude:

\begin{proposition}
  If $\langle e_i |i \in \mathbbm{N} \rangle$ enumerates those indices of
  Turing programs that ``construct'' in the above sense hyperarithmetic sets
  $S_{e_i}$, {\tmem{via}} recursive trees, we may define a single MH
  `hyperarithmetically deciding'', HYPD, spacetime region in which any query
  of the form $? n \in S_{e_i} ?$ can be answered in finite time.
\end{proposition}

In the above, we may define the HYPD region by piecing together parts of
spacetime that are ``$S_{e_1}$-deciding'' just as an AD region can be so
defined: an overall control machine can take $\langle i, n \rangle$ as input
and activate the $i$'th machine constructing $S_{e_i}$. It is worth
emphasising that no machine in this tree array is itself performing
``supertasks'' ({\tmem{i.e.}} performing infinitely many actions in its own
proper time), but if it issues a signal to another process, it does so only
once after a finite amount of its own proper time. However a tree structure
for such a region can no longer have recursive ordinal height, as it can be
shown (again see {\cite{Rog}}) that the ranks of finite path trees
constructing hyperarithmetic sets exhaust all the recursive ordinals. \ Thus a
MH spacetime may be able to calculate ``beyond'' any hyperarithmetic set, but
the spacetime structure is not realised by a recursive tree. Moreover as noted
above, {\tmem{any}} finite path tree can be realised in a MH spacetime: thus
there apparently is no {\tmem{a priori}} upper bound on computational
complexity.

\subsection{The complexity of questions decidable in Kerr spacetimes }

Etesi and N\'emeti consider a special class of MH spacetimes, Kerr spacetimes,
which contain rotating blackholes. (The reasons, which we shall not discuss,
of why they choose this particular spacetime, is that the physics of
communication between the observer moving along an infinite $\gamma_1$ to that
on $\gamma_2$ is less problematic.) Their Proposition 1 shows, just as
{\cite{Ho92}} did, that any $\Sigma_1$ predicate can be decided in their
arrangement. They also obey Assumption 2 that there is no swamping, and in
fact, initially at least, the observer $\mathcal{O}_r$ travelling along
$\gamma_1$ only sends a single signal to $p$ on $\gamma_2$. They further
remark though (Prop. 2) that in fact the set-up allows for resolving of
queries $? n \in R ?$ for sets slightly more complicated: $R$ can be taken as
a union of a $\Sigma_1$ and a $\Pi_1$ set - and thus need not be in either
class. Indeed they indicate an argument at Prop.3, that if the observer
$\mathcal{O}_r$ is allowed to send $k$ different signals, (they take $k = 2)$
then any $k$-fold Boolean combination of $\Sigma_1$ and $\Pi_1$ sets $R =
\bigcap_{i < k - 1} (S^i \cup P^i)$ \ (with $S^i \in \Sigma_1$ and $P^i \in
\Pi_1$) can be decided. They ask how far in the arithmetical hierarchy this
can kind of argument can be taken.

The classes of predicates just indicated all fall within $\Delta_2$
($=_{\tmop{df}} \Sigma_2 \cap \Pi_2)$. However, even allowing $k$ to go to
$\infty$ will not exhaust this class. The class of $\Delta_2$ predicates can
be characterised using Turing machines that may write to an output tape a
single 0 or 1 digit, which the machine may however later change (this class
was studied, and so characterised, by a number of persons, Davis, Gold
{\cite{Go65}}, Putnam {\cite{Pu65}}).

\begin{definition}
  (Putnam {\cite{Pu65}}) $R \subset \mathbbm{N}$ is {\tmem{a trial and error
  predicate}} if there are Turing machines $M_0, M_1$ of this sort so that
  
  (i) each of $M_{0,} M_1$ changes its mind about its output at most finitely
  many times, for any input $n$, and
  
  (ii) \ $n \in R \Longleftrightarrow$the eventual value of $M_0$'s output
  tape on input $n$ is 1
  
  \ \ \ \ \ $n \notin R \Longleftrightarrow$the eventual value of $M_1$'s
  output tape on input $n$ is 0.
\end{definition}

We then have that $R \subset \mathbbm{N}$ is $\Delta_2$ if and only if it is
a trial and error predicate. Can Etesi and N\'emeti's analysis of Boolean
combinations of universal and co-universal predicated be extended to trial and
error predicates? The answer has to be no, if the machinery is required to
only send a fixed number of signals to $p$ on $\gamma_2$: the observer $r$ on
$\gamma_1$ may now run two Turing machines $M_0, M_1$ of the above type, and
may send a signal $S^i$ to $p$ \ each time $M_i$ changes its output digit, but
crucially we do not know in advance how many times that will be. It is not
that $p$ {\tmem{will}} or {\tmem{may}} receive infinitely many signals (so it
is not ``swamping'' that is the problem), but that $p$ will have to be
prepared to receive a {\tmem{potential}} infinity of signals of one type, if
the arrangement is to resolve $\Delta_2$ predicates. It is not hard to show
that there are $\Delta_2$ predicates $R$ so that for any representation in
terms of machines of the type $M_i$ above, there will be no recursive
functions $f_i$ so that $f_i (n)$ bounds the number of times that $M_i$
changes its mind about input $n$. (Hence for deciding $\Delta_2$-predicates,
it is not just that we cannot do this if there is a fixed $k$ on the number of
alternations which works for any $n$ as input; we cannot, given the input $n$
run some other initial recursive test on $n$ to determine {\tmem{in advance}}
what the $k = k (n)$ should be for this particular $n$.

However if we relax this fixed bounding assumption on the number of signals
that can be transmitted to $p$ (but still require the number to be finite so
that Assumption 2 holds) then both these spacetimes (and also $\tmop{SAD}_1$
spacetimes) can compute $\Delta_2$ (but not in general $\Sigma_2$) predicates.
We summarise this discussion as:

{\tmstrong{Theorem B.}} {\tmem{The relations $R \subseteq \mathbbm{N}$
computable in such spacetimes form a subclass of the $\Delta_2$-relations on
$\mathbbm{N}$; this is a proper subclass if and only if there is a fixed
finite bound on the number of signals sent to the observer on the finite
length path.}}

\section{An upper bound on computational complexity for each MH spacetime}

There are only countably many hyperarithmetic sets (as there are only
countably many recursive trees) whilst there are uncountably many
non-recursive finite path trees. \

However Hogarth's construction of AD spacetime regions, which we have
extended to hyperarithmetic deciding regions, only depends on being able to
take the ``toy'' spacetime construction at the beginning and reproduce that in
a suffiently nested manner in order to create more and more complicated
regions. It is possible that a piece of an MH spacetime has a region that
reflects the structure of a finite path tree that is {\tmem{not}} recursive:
given {\tmem{any}} finite path tree $T$ one can ``construct'' an MH spacetime
$\mathcal{M}_T$ satisfying the requirement that its $\tmop{SAD}$ components
reflect the structure of $T$. The reader may very well have several objections
at this point (if not well before): (i) that in general we may have no way of
conceiving of a \ ``problem'' or ``calculation'' \ at an arbitrary level of
complexity in this hierarchy; that (ii) if we could we would not be able to
initiate, or `` control'' the hardware for the problem (or even locate the
relevant spacetime region in which it could be performed!).

Just to consider the objection (ii) first: I think this problem already arises
for an AD spacetime: this contains an infinite string of Turing machines, that
whilst acting independently to some extent in their own patches of spacetime
$O_j$, and not swamping each other with messages, cannot conceivably be ``set
up'' or initialised by finite beings in a finite time, to perform the task in
our mind. Hogarth imagines the machinery all set up and ready to go: it is
just waiting for our input and the switch to be thrown. However we can imagine
that here too. For (i): we can readily conceive of, or think up, mathematical
questions about $\Sigma_{\omega}$ or $\Sigma_{\omega + \omega}$ or
$\Sigma_{\omega^2} \ldots$ for larger recursive ordinals the questions are
more likely to be intimately related to those ordinals, rather than some
general problem in number theory. Indeed it is not clear anyway what the class
of hyperarithmetic sets is: for example, we are unable to enumerate them in
any effective way: we may define them as those sets built up using those
Turing machines $P_{e_i}$ which output codes of finite path trees, but in fact
we have no effective or algorithmic way of knowing of a given index $e$
whether $P_e $ does or does not code a finite path tree. In general whether
$P_e$ codes a (not necessarily finite) path tree is iself a $\Pi_1$ property
of $e$ but that it codes a {\tmem{finite}}, that is, a {\tmem{wellfounded}}
path tree is far even from arithmetic{\footnote{The set of indices coding
finite path trees is a complete $\Pi^1_1$-set of integers: it thus requires a
universal function quantification in analysis, or second order number theory
(see for example, {\cite{Rog}} Thm.{\tmem{XX}}.)}}. Supposing we did actually
have a part of our spacetime which was HYPD (and we could recognise it!) then
given an integer $n$ we would not know in full and absolute generality, which
indices $e$ it is suitable to even ask $? n \in S_e ?$ \ However this is not
really the point of the argument: we are looking at logico-mathematical
boundaries to these kinds of spacetimes (and not anthropomorphic limitations).

Any set of integers $B \subseteq \mathbbm{N}$ whatsoever can be seen as
resolvable by a $\tmop{SAD}_{}$ spacetime region containing a finite path tree
structure $T$ with an assignment of basic open sets, {\tmem{i.e.}} recursive
sets, to the terminal nodes: it is that we can cheat and hardwire the answers
to queries concerning $B$ {\tmem{into}} the very structure of the tree $T$
beforehand (just as any set $B$ indeed has a code according to the definition
above, but again for trivial reasons).

At least we shall not have uncountably many worries of this sort as the
following arguments show.

\begin{definition}
  Let $\mathcal{M}= (M, g_{\tmop{ab}})$ be a spacetime. We define $w
  (\mathcal{M})$ to be the least ordinal $\eta$ so that $\mathcal{M}$ contains
  no $\tmop{SAD}$ region whose underlying tree structure has ordinal rank
  $\eta$.
\end{definition}

Note that $0 \leq w (\mathcal{M}) \leq \omega_1$ ($0 = w (\mathcal{M})$
implies that $\mathcal{M}$ contains no SAD regions whatsoever, that is, is not
MH; the upper bound is for the trivial reason that every finite path tree is a
countable object and so cannot have uncountable ordinal rank).

\begin{proposition}
  For any spacetime $\mathcal{M},$ w($\mathcal{M}) < \omega_1$.
\end{proposition}

Proof: We here use our assumptions on our manifolds: Assumption (i) that for
different $\eta$ the different $\tmop{SAD}_{\eta}$ component occupy disjoint
open regions $O_{\eta}$ of the manifold, and (ii) that the manifold is
separable (which follows from paracompactness and being Hausdorff). Let $X
\subset M$ be a countable dense subset of $M$. Then each open region
$O_{\eta}$ of $M$ contains members of $X$. As disjoint regions contain
differing members of $X$ there can only be countably many such regions
$O_{\eta} \subset M,$ and therefore a countable bound. \ \ \ \ \ \ \ \ \ \ \ \
\ \ \ \ \ \ \ \ \ \ \ QED

(Of course this is only the generalisation of the argument that in
$\mathbbm{R}$ there can only be countably many disjoint open intervals of the
form $(a, b)$!). This Proposition is thus our Theorem B

Consequently, if {\tmem{our}} spacetime $\mathcal{M}_0$ happens to be MH, then
for some countable universal constant of our universe, $w (\mathcal{M}_0)$,
the possible ``calculations'' theoretically performable all have complexity
bounded by $\text{$w (\mathcal{M}_0)$}$. Of course it may well be that the
reader (like this author) believes instead that $\text{$w (\mathcal{M}_0)$} =
0!$

\end{document}